\documentclass{elsart}

\usepackage{psfig}

\def\gtrsim{\mathrel{\hbox{\rlap{\hbox{\lower4pt\hbox{$\sim$}}}\hbox{$>$}}}}
\def\lesssim{\mathrel{\hbox{\rlap{\hbox{\lower4pt\hbox{$\sim$}}}\hbox{$<$}}}}
\def\fs{\hbox{$.\!\!^{\rm s}$}}
\def\farcs{\hbox{$.\!\!^{\prime\prime}$}}

\def\astrobj#1{#1}

\begin{document}

\begin{frontmatter}

\title{The extragalactic nature of the serendipitous \\
{\it BeppoSAX} source \astrobj{2MASX 
J14585116$-$1652223}\thanksref{labeltitle}}
\thanks[labeltitle]
{Based on observations collected at the Astronomical Observatory of 
Bologna in Loiano, Italy.}

\author[add1]{N. Masetti\corauthref{cor1}},
\author[add2]{E. Mason},
\author[add1]{L. Bassani},
\author[add1]{R. Landi},
\author[add1,add3]{E. Maiorano},
\author[add1]{A. Malizia}
\author[add1]{and E. Palazzi}

\corauth[cor1]{corresponding author; email: \texttt{masetti@bo.iasf.cnr.it}}

\address[add1]{INAF -- Istituto di Astrofisica Spaziale e Fisica Cosmica, 
Sezione di Bologna, Via Gobetti 101, I-40129 Bologna, Italy (formerly 
IASF/CNR, Bologna)}

\address[add2]{European Southern Observatory, Casilla 19001, Santiago 19, 
Chile}

\address[add3]{Dipartimento di Astronomia, Universit\`a di Bologna, Via 
Ranzani 1, I-40126 Bologna, Italy}

\date{. Received 21 July 2005 / Accepted 28 September 2005}

\begin{abstract} 
As a follow-up of the X--ray serendipitous detection of the
source \astrobj{2MASX J14585116$-$1652223}, medium-resolution optical 
spectroscopic
observations collected at the 1.5-metre ``Cassini" telescope of the
Astronomical Observatory of Bologna were performed. This allowed us to
determine the extragalactic nature of this X--ray source, which is a Type
2 Seyfert galaxy at redshift $z$ = 0.068 $\pm$ 0.001. 
This result points to the fact that \astrobj{2MASX J14585116$-$1652223}
hosts one of the very few Active Galactic Nuclei detected so far in the 
hard X--ray band above 20 keV, in particular at $z >$ 0.05. Other 
optical and X--ray properties of this object are also discussed.
\end{abstract}


\end{frontmatter}



\section{Introduction}

The X--ray emitting object \astrobj{2MASX J14585116$-$1652223} was 
serendipitously
discovered by {\it BeppoSAX} and detected in the 1.5--100 keV range during
the observation of the Seyfert 2 NGC 5793. Details on the discovery are
reported in Landi et al. (2005). 
The source, located at coordinates (J2000) $\alpha$ = 14$^{\rm h}$
58$^{\rm m}$ 50$\fs$7, $\delta$ = $-$16$^{\circ}$ 51$'$ 50$\farcs$7 
is, according to the NASA/IPAC Extragalactic Database\footnote{available at 
{\tt http://nedwww.ipac.caltech.edu/}}, a bright irregular spiral galaxy 
belonging to the Two-Micron All Sky Survey eXtended (2MASX) Source 
Catalog\footnote{available at {\tt
http://www.ipac.caltech.edu/2mass/}} (Skrutskie et al. 1997).
A DSS-II-Red\footnote{available at {\tt http://archive.eso.org/dss/dss/}} 
image of the optical field of this source is shown in Fig. 1.

The source was detected both by MECS and PDS intruments onboard {\it
BeppoSAX} and showed a spectral shape described by a flat power law (of
photon index $\Gamma \sim 1.2$) with a 2--10 and 20--100 keV flux of 
$\sim$10$^{-12}$ erg cm$^{-2}$ s$^{-1}$ and $\sim$10$^{-11}$ erg cm$^{-2}$
s$^{-1}$, respectively. These spectral characteristics imply that 
\astrobj{2MASX
J14585116$-$1652223} is an Active Galactic Nucleus (AGN) which is either
absorbed at low energies (by an hydrogen column $N_{\rm H}$ 
$\sim$ 4$\times$10$^{22}$ cm$^{-2}$) or with an intrinsically flat 
spectrum. Due to the lack of spectral information below 1.5 keV it 
was not possible to tell which of the two interpretations above holds 
true for the source.

Very little is known about this source except for its
optical/near-infrared (NIR) magnitudes. The source is fairly bright with a
total 2MASX NIR photometry of 14.4, 13.6 and 13.0 magnitudes in the $J$,
$H$ and $K$ bands, respectively, and an extension of 11$''$. The optical
counterpart has magnitudes $B \sim$ 16.4 and $R \sim$ 15.1 according to
the USNO-A2.0 catalog\footnote{available at {\tt
http://archive.eso.org/skycat/servers/usnoa/}}. No X--ray detections other 
than the {\it BeppoSAX} one are reported in the literature. The source was 
however observed by {\it ROSAT} in the 0.1--2.4 keV band, but no detection 
was achieved; this latter fact is again consistent with the presence of 
strong absorption. No detection in the radio bands was found in the 
literature.

In order to confirm and classify more accurately the nature of this 
source, we performed medium-resolution optical spectroscopic observations 
on it with the 1.5-metre telescope of the Astronomical Observatory of 
Bologna. This paper is organized as follows: Sect. 2 describes the 
observations, Sect. 3 will present the results and a discussion on them, 
whereas in Sect. 4 the conclusions are drawn.

\section{Optical observations}
                                                                                
The Bologna Astronomical Observatory $1.52$~metre ``G.D. Cassini'' 
telescope plus BFOSC was used to spectroscopically observe \astrobj{2MASX 
J14585116$-$1652223}. The pointing was performed on June 6, 2005, under 
non-optimal weather conditions (passing thick clouds; the seeing was 
$\sim$2$''$). The observation was made of two consecutive 
spectroscopic integrations of equal exposure time; the details of these 
spectroscopic observations are reported in Table 1. The BFOSC 
instrument is equipped with a $1300\times1340$ pixels EEV CCD. Grism \#4 
and a slit width of $2''$ were used, providing a nominal spectral coverage 
of the 3500-8500 \AA~range and a final dispersion of $4.0$~\AA/pix.

Spectra, after cosmic-ray rejection, were reduced in the usual fashion;
they were then background subtracted and optimally extracted (Horne 1986)  
using IRAF\footnote{IRAF is the Image Analysis and Reduction Facility made
available to the astronomical community by the National Optical Astronomy
Observatories, which are operated by AURA, Inc., under contract with the
U.S. National Science Foundation. It is available at {\tt
http://iraf.noao.edu/}}. Wavelength calibration was performed using He-Ar
lamps; the spectra were then flux-calibrated by using the
spectrophotometric standard \astrobj{BD+25$^\circ$3941} (Stone 1977). The 
two spectra were eventually stacked together to increase the S/N ratio. 
Wavelength calibration was cross-checked using the night sky lines; 
the error is $\sim$0.5~\AA.

\section{Results and discussion}

The spectrum of \astrobj{2MASX J14585116$-$1652223} acquired in Loiano 
(Fig. 2), albeit noisy, shows a number of emission features which we 
identified with redshifted optical lines typical of active galaxies. 
These include [O~{\sc iii}] $\lambda\lambda$4958,5007, H$_\alpha$ and 
[N~{\sc ii}] $\lambda\lambda$6549,6583. All identified emission lines 
are reported in Table 2 and yield a redshift of $z$ = 0.068 $\pm$ 0.001. 
Thus, \astrobj{2MASX J14585116$-$1652223} is indeed an extragalactic 
object.
Besides the extragalactic nature of this source, we also emphasize 
that this result, together with the X--ray information gathered with {\it 
BeppoSAX} (Landi et al. 2005), indicates that \astrobj{2MASX 
J14585116$-$1652223} is the host galaxy of one of the few hard X--ray AGNs 
detected so far in the 20--100 keV band. In particular, this is one of the 
very few AGNs detected in the hard X--ray band and lying at redshifts 
larger than $z$ = 0.05 (e.g., Perola et al. 2002; Risaliti 2002).

The fluxes of the emission features observed in the spectrum of Fig. 2 are 
reported as well in Table 2 and are corrected for the Galactic absorption. 
This correction was applied assuming a color excess $E(B-V)$ = 0.10 
towards the direction of \astrobj{2MASX J14585116$-$1652223} (Schlegel et 
al. 1998) and the Galactic extinction law of Cardelli et al. (1989). 
The H$_\beta$ emission is not detected: the 3$\sigma$ upper limit 
of its flux is reported in Table 2.

The relative narrowness ($\approx$800 km s$^{-1}$) of the H$_\alpha$ line 
width is found to be comparable with those of forbidden nebular lines of 
[O~{\sc iii}] and [N~{\sc ii}]; moreover, no broad component for the 
Balmer lines is detected. The observed emission line flux ratios 
[N~{\sc ii}]/H$_\alpha$ $\sim$ 1 and [O~{\sc iii}]/H$_\beta$ $>$ 10 
clearly identify this source as a Type 2 Seyfert galaxy when compared to
the diagnostic diagrams of Veilleux \& Osterbrock (1987).
This classification fits with the suggestions of Landi 
et al. (2005) who, through the analysis of X--ray data from this source 
collected with {\it BeppoSAX}, conclude that this is an AGN with substantial 
intrinsic absorption in the X--ray band.

The lower limit on the H$_\alpha$/H$_\beta$ line ratio can be used to 
give an estimate of the lower limit to the extinction towards the 
narrow-line region (NLR) of the AGN hosted by the galaxy \astrobj{2MASX 
J14585116$-$1652223}. The H$_\alpha$/H$_\beta$ flux ratio, once corrected 
for Galactic absorption, is $>$6.2. Next, considering an intrinsic Balmer 
decrement of H$_\alpha$/H$_\beta$ = 3.1 for the gas conditions 
in a NLR (Osterbrock 1989) and the 
extinction law by Cardelli et al. (1989), the lower limit on the 
Balmer decrement H$_\alpha$/H$_\beta$ $>$ 6.2 implies an extinction $A_V$ 
$>$ 2.2 mag (in the galaxy rest frame).
This latter lower limit can be compared with the neutral hydrogen column 
density $N_{\rm H}$ inferred from the X--ray spectrum of the source 
obtained with {\it BeppoSAX}. Using the empirical formula of Predehl \& 
Schmitt (1995), $N_{\rm H}$ = $A_V$(mag) $\cdot$ 1.79$\times$10$^{21}$ 
cm$^{-2}$, the lower limit on $A_V$ implies a hydrogen column density 
towards the NLR of this AGN of $N_{\rm H} >$ 3.9$\times$10$^{21}$ 
cm$^{-2}$. This limit is fully consistent with the value for $N_{\rm H}$ 
inferred from X--ray data ($N_{\rm H}$ = 4$\times$10$^{22}$ cm$^{-2}$).

The comparison between the [O~{\sc iii}] $\lambda$5007 emission flux (see
Table 2) and the 2--10 keV X--ray flux implies an X--ray/[O~{\sc
iii}]$_{\rm 5007}$ ratio of $\sim$80, indicating that the source is in the
Compton-thin regime (see Bassani et al. 1999). We however caution the
reader that this is an upper limit to the ratio, since the 
flux of the [O~{\sc iii}] $\lambda$5007 line is not corrected for 
dust extinction intrinsic to the galaxy \astrobj{2MASX 
J14585116$-$1652223}.

Using a cosmology with $H_{\rm 0}$ = 71 km s$^{-1}$ Mpc$^{-1}$,
$\Omega_{\Lambda}$ = 0.7 and $\Omega_{\rm m}$ = 0.3, we find that the
luminosity distance to the galaxy \astrobj{2MASX J14585116$-$1652223} is 
$d_L$ = 307 Mpc, and that its X--ray luminosities are 1.0$\times$10$^{43}$~erg
s$^{-1}$ and 1.2$\times$10$^{44}$~erg s$^{-1}$ in the 2--10 keV and
20--100 keV bands, respectively. In parallel, again by correcting for the
Galactic reddening, the absolute $B$-band magnitude of the galaxy is $M_B
\sim$ $-$21.4; we again remark that the latter is actually a lower limit
to the optical luminosity, as no correction for absorption intrinsic to
the object was made. These values indicate that this source is a fairly
luminous Type 2 AGN (see, e.g., Malizia et al. 1999).

\section{Conclusions}

Through medium-resolution optical spectroscopy collected at the Cassini
telescope of the Astronomical Observatory of Bologna we determined the
nature and the distance of the serendipitous source \astrobj{2MASX
J14585116$-$1652223}, formerly detected in X--rays with {\it BeppoSAX}. We
identified this object as a Type 2 Seyfert galaxy at redshift $z$ = 0.068
$\pm$ 0.001. This redshift implies a luminosity distance of $d_L$ = 307
Mpc to the source, which in turn allowed us to state that this AGN is in
the bright side of the distribution both in X--rays and in optical. 
This result also indicates that this object is the host galaxy of one 
of the very few AGNs detected up to now in the hard X--ray band between 20 
and 100 keV, especially at redshifts larger than $z$ = 0.05.
The acquired spectrum also gave us the possibility of determining other
parameters for this galaxy such as its Compton-thinness and a lower 
limit on the reddening towards the NLR of the AGN hosted in the galaxy
\astrobj{2MASX J14585116$-$1652223}.

In the close future, thanks to their capabilities, {\it INTEGRAL} and 
{\it ASTRO-E2} (``{\it Suzaku}") will provide larger samples of distant 
sources detected at energies above 20 keV, to which the properties of 
\astrobj{2MASX J14585116$-$1652223} will be compared.

{\em Acknowledgements.}
We thank A. De Blasi for the night assistance in Loiano.
This research has made use of the NASA Astrophysics
Data System Abstract Service, of the NASA/IPAC Extragalactic Database
(NED), and of the NASA/IPAC Infrared Science Archive, which are operated
by the Jet Propulsion Laboratory, California Institute of Technology,
under contract with the National Aeronautics and Space Administration.
This research has also made use of the SIMBAD database operated at CDS,
Strasbourg, France. NM thanks the European Southern Observatory in
Vitacura, Santiago (Chile) for the pleasant hospitality during the
preparation of this paper. We moreover thank the anonymous referee for 
several useful remarks and comments which helped us to improve this paper.

{}

\clearpage

\begin{table}
\caption[]{Log of the spectroscopic observations of \astrobj{2MASX
J14585116$-$1652223} presented in this paper.}
\begin{center}
\begin{tabular}{lcccc}
\noalign{\smallskip}
\hline
\noalign{\smallskip}
\multicolumn{1}{c}{Date} & Mid-exp. & Grism & Slit width & Exp. \\
 & time (UT) & & (arcsec) & time (s) \\
\noalign{\smallskip}
\hline
\noalign{\smallskip}

06 Jun 2005 & 20:59:04 & \#4 & 2.0 & 2$\times$1800 \\
                                                                                
\noalign{\smallskip}
\hline
\noalign{\smallskip}
\end{tabular}
\end{center}
\end{table}

\begin{table}[t!]
\caption[]{Observer's frame wavelengths (in \AA ngstroms) and fluxes (in 
units of 10$^{-15}$ erg s$^{-1}$ cm$^{-2}$) of the emission lines detected 
in the spectrum of Fig. 2. Flux values are corrected for Galactic absorption 
assuming $E(B-V)$ = 0.10 along the \astrobj{2MASX J14585116$-$1652223} 
line of sight (Schlegel et al. 1998). The error on the line positions is 
conservatively assumed to be $\pm$4 \AA, i.e., comparable with the spectral 
dispersion (see text). The upper limit on the flux of the H$_\beta$ 
emission line was computed at the expected wavelength for this line 
(reported in square brackets) assuming a redshift $z$ = 0.068 for the 
source.}
\begin{center}
\begin{tabular}{clr}
\noalign{\smallskip}
\hline
\noalign{\smallskip}
$\lambda_{\rm obs}$ (\AA) & \multicolumn{1}{c}{Line} & 
\multicolumn{1}{c}{Flux} \\
\noalign{\smallskip}
\hline
\noalign{\smallskip}

 [5192] & H$_\beta$            &   $<$0.9 \\
  5296  & $[$O {\sc iii}$]$ $\lambda$4958 & 2.4$\pm$0.5 \\
  5349  & $[$O {\sc iii}$]$ $\lambda$5007 & 9.0$\pm$0.9 \\
  6993  & $[$N {\sc ii}$]$ $\lambda$6549  & 2.3$\pm$0.4 \\
  7012  & H$_\alpha$                      & 5.6$\pm$0.6 \\
  7035  & $[$N {\sc ii}$]$ $\lambda$6583  & 6.3$\pm$0.6 \\

\noalign{\smallskip}
\hline
\noalign{\smallskip}
\end{tabular}
\end{center}
\end{table}

\clearpage

\begin{figure}[t!]
\begin{center}
\psfig{file=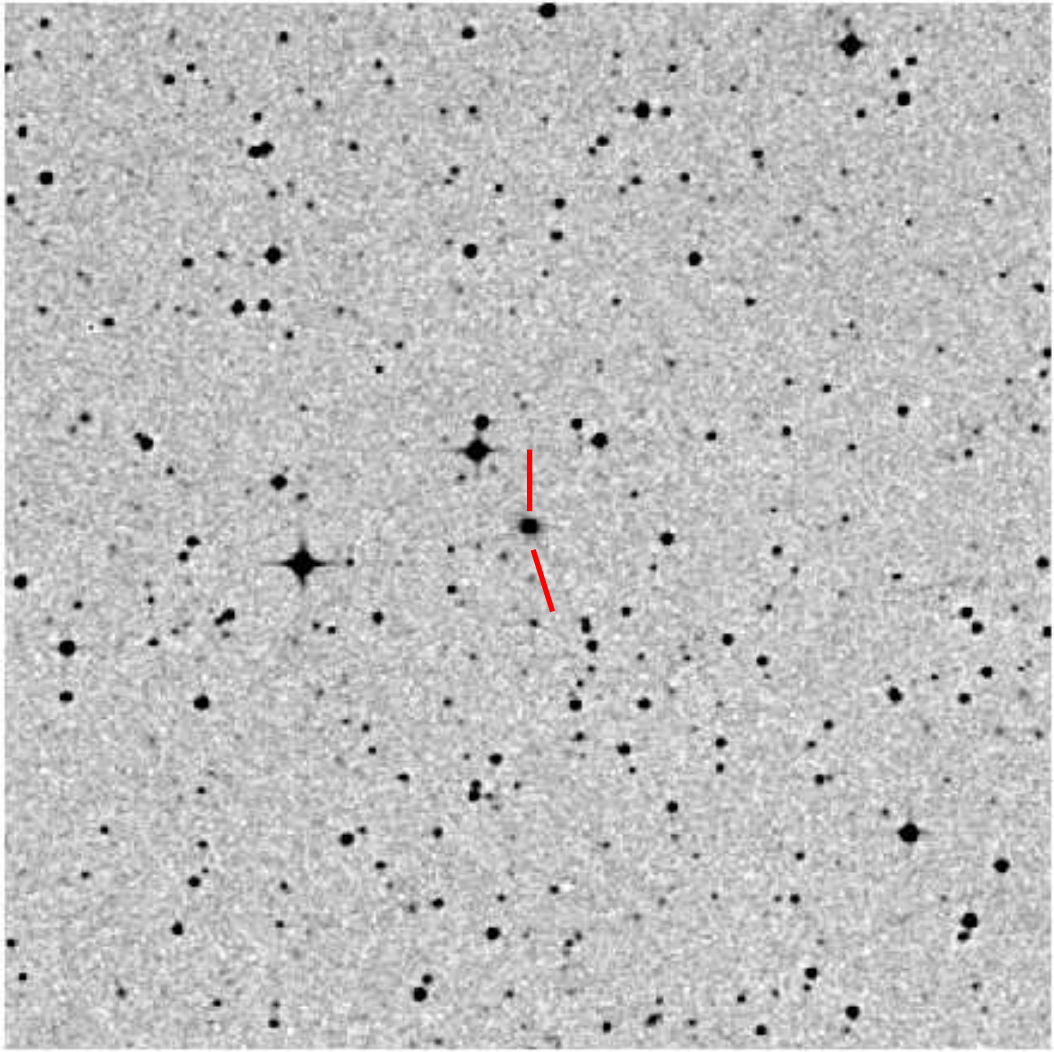,width=14cm}
\end{center}
\caption{DSS-II-Red image of the \astrobj{2MASX J14585116$-$1652223} 
field. North is
up and East to the left. The field size is 10$'$$\times$10$'$. The
tick marks indicate the position of the source.}
\end{figure}

\clearpage

\begin{figure}[t!]
\hspace{-2cm}
\psfig{file=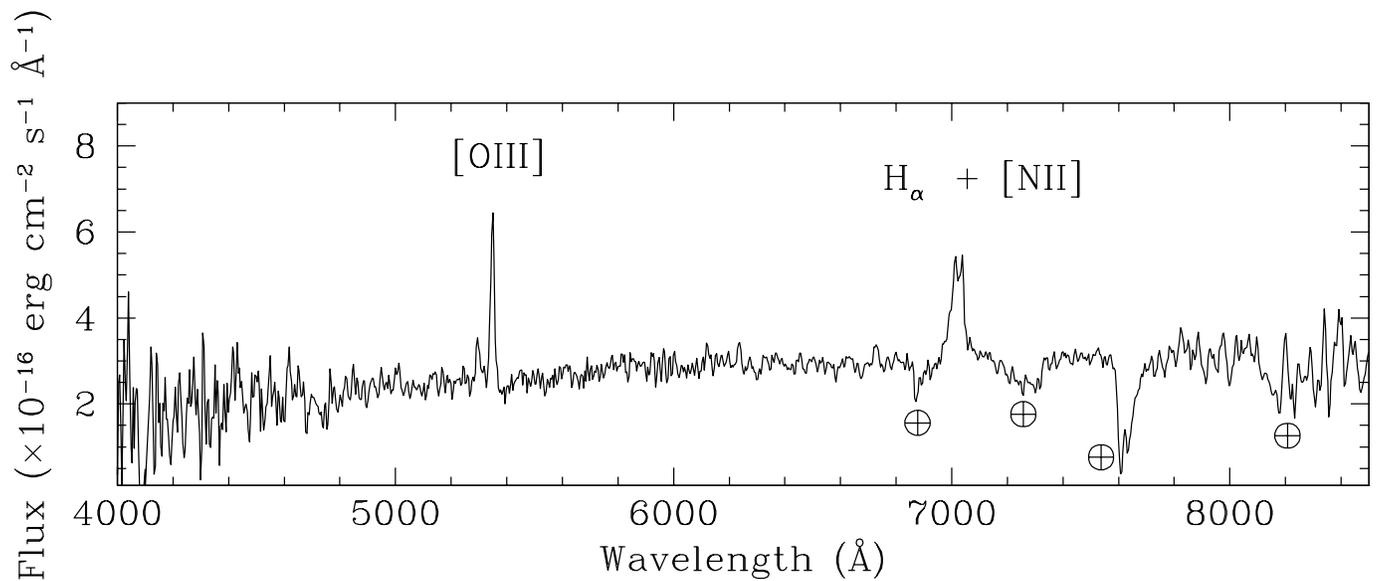,width=18.8cm,angle=270}
\caption{Average optical spectrum (not corrected for the Galactic
reddening) of \astrobj{2MASX J14585116$-$1652223} acquired with the 
Cassini telescope in Loiano. The main spectral features are labeled 
(see also Table 2). These allowed us to determine the redshift of the 
source as $z$ = 0.068. The symbol $\oplus$ indicates atmospheric 
telluric features.}
\end{figure}

\end{document}